\documentclass[12pt,preprint]{aastex}

\begin{document}

\def\pdot {\dot P}
\def\Omdot {\dot \Omega}
\def\ltsima{$\; \buildrel < \over \sim \;$}
\def\lsim{\lower.5ex\hbox{\ltsima}}
\def\gtsima{$\; \buildrel > \over \sim \;$}
\def\gsim{\lower.5ex\hbox{\gtsima}}
\def\msole{~M_{\odot}}
\def\mdot {\dot M}
\def\ee {1E~1207--5209~}
\def\cha {\textit{Chandra~}}
\def\psr{PSR J0737--3039~}
\def\xmm  {\textit{XMM-Newton~}}

\title{A first XMM-Newton look at the relativistic double pulsar
PSR J0737-3039}

\author{A. Pellizzoni, A. De Luca, S. Mereghetti, A. Tiengo, F. Mattana, P. Caraveo}
\affil{Istituto di Astrofisica Spaziale e Fisica Cosmica, \\
Sezione di Milano  ``G.Occhialini'' - CNR \\
v.Bassini 15, I-20133 Milano, Italy \\
alberto@mi.iasf.cnr.it}
\author{M. Tavani}
\affil{Istituto di Astrofisica Spaziale e Fisica Cosmica, \\
Sezione di Roma - CNR \\
v.Fosso del Cavaliere 100, I-00133 Roma, Italy \\}
\author{G.F. Bignami}
\affil{Centre d'\'Etude Spatiale des Rayonnements, CNRS-UPS, \\
9, avenue du Colonel Roche, 31028 Toulouse Cedex 4, France \\
and Universit\`a degli Studi di Pavia, Dipartimento di Fisica 
Nucleare e Teorica, \\
v.Bassi 6, I-27100 Pavia, Italy}

\begin{abstract}
We present the results of a 50 ks long X--ray observation of the
relativistic double pulsar system \psr obtained with the \xmm
satellite in March 2004. The source has a soft spectrum (power law
photon index = $3.5 ^{+0.5} _{-0.3}$     ) and a 0.2-10 keV 
luminosity of $\sim3 \times 10^{30}~erg~s^{-1}$ (assuming a distance 
of 500 pc), consistent with the values derived from a previous \cha
observation. No flux variations have been detected, implying the
absence of large orbital  modulations.
The high time resolution of the EPIC instrument has allowed us to
perform the first search for X--ray pulsations from this system.
The result was negative, with an upper limit of 60\% on the pulsed
fraction of the 22 ms pulsar.
\end{abstract}


\section{Introduction}

The short-period double radio pulsar system \psr (Burgay et al.
2003; Lyne et al. 2004) is of paramount interest as a probe 
for theories of strong field gravity and represents a unique 
laboratory for the study of the magnetospheres of radio pulsars.
The system consists of a fast, recycled radio pulsar (PSR A,
P=22.7 ms) orbiting its slower companion (PSR B, P=2.77 s) with
an orbital period of only 2.4 hours. The fact that we observe it
nearly edge-on (inclination $i=87^{\circ}$), provides an
additional tool to further constrain the magnetospheric structures
by reciprocal occultation of the two pulsars and their shocked
winds near conjunction.

Due to the interaction of the relativistic wind of the fast
spinning pulsar A ($\dot{E}^{A}_{rot}$=5.8$\times10^{33}$ erg
s$^{-1}$) with the magnetosphere of its much less energetic
companion ($\dot{E}^{B}_{rot}$=1.6$\times10^{30}$ erg s$^{-1}$),
the formation of a bow-shock, likely  emitting at high energies,
is expected. The predicted fluxes are roughly comparable to those
expected for  magnetospheric and/or surface emission from the
pulsars. Alternately, most of the high-energy emission could arise
from the interactions between the pulsar relativistic winds and
the interstellar medium. Detailed emission models and geometry are
clearly speculative without valuable high-energy observations,
particularly at X--ray energies.

\psr was first detected in the 0.2-10 keV range during  a short
(10 ks) \cha observation (McLaughlin et al. 2004), with an X-ray
luminosity of L$_{x}=2\times10^{30} (d/0.5~kpc)^{2}$ erg s$^{-1}$, roughly
equal to the entire spin-down luminosity of the slow pulsar B and
corresponding to only a small fraction of the spin-down luminosity
of pulsar A. With the detection of $\sim$80 pulsar photons
spanning one orbital period, the \cha data could only poorly
constrain the source spectrum and were of limited use for the
study of variability. Furthermore the time resolution of the ACIS
instrument was not enough to search for pulsations at the spin
periods of the two pulsars.

In this letter we present the analysis of a 50 ks long \xmm
observation of \psr  performed through the Director's
Discretionary Time program and made publicly available. Although
episodes of high particle background affected this observation,
these data yield an improvement in  statistics by a factor
$\sim$10 and have allowed us to carry out the first search for
fast pulsations from this system. In addition, the coverage of
$\sim$5 orbits allows to set limits on the orbital variability
of the X--ray emission.

\section{XMM-Newton observation and data reduction}

The \xmm observation started on 2004, March 28 at 07:47:26 UT
and lasted $\sim$51 ks. We report here on data collected
by the EPIC instrument, which consists of a pn CCD camera
(Str\"uder et al. 2001) and of two MOS CCD detectors
(Turner et al. 2001), for a total total collecting area of
$\sim$2500 cm$^2$ at 1.5 keV. The pn camera was operated in
``Fast Timing'' mode, with a time resolution of 0.03 ms,
achieved at the cost of the loss of positional information
along the detectors columns and of a higher background. The
MOS1 camera was operated in ``Full Frame'' mode (2.6 s time
resolution, imaging across the full 15$'$ radius telescope
field of view), while the MOS2 was set in ``Small Window''
mode (0.3 s time resolution, imaging on a $\sim1' \times1'$
portion of the central CCD\footnote{Peripheral CCDs are read
with a 2.6 s frame time as in Full Frame mode}). The Observation
Data Files (ODFs) were retrieved through the \xmm Science
archive. The data reduction was performed using the most recent
release of the \xmm Science Analysis software (SASv6.0.0),
with the standard pipeline tasks ({\em epproc} for pn data and
{\em emproc} for MOS data).
\psr is a faint X-ray source with a flux of a few 10$^{-14}$
erg cm$^{-2}$ s$^{-1}$ and therefore a careful study of the
instrumental background is required. Two different issues are
of particular relevance:

\begin{itemize}

\item {\em Soft flares} in pn data. These are due to the
interaction of high-energy particles with the CCD; they are
seen as short (0.1-0.5 s) and intense (up to thousands of counts
s$^{-1}$) bursts of events, with a typical energy of $\sim0.22$
keV (mono-pixel events) and $\sim0.45$ keV (bi-pixel events) and
with a highly anisotropic distribution across the CCD, which
prevents standard subtraction. Following Burwitz et al. (2004),
we used an {\em ad hoc} Good Time Interval (GTI) filter  to
remove such background component. This was done by intersecting
two independent GTI files which were built by setting a
threshold of 15 cts s$^{-1}$ on boxcar-smoothed light curves
(0.1 s time bins) extracted in the 0.2-0.3 keV range for
mono-pixel events and in the 0.4-0.5 keV range  for bi-pixel events.
The rejected frames amount to $\sim5400$ s ($\sim$11\% of the
observing time). As a result, the soft flare background
component is almost totally suppressed.

\item {\em Soft proton flares}. The whole observation is affected
by medium-intensity proton flares. This is a point of concern
especially for the pn camera, since in Fast Timing mode the
collapse of data along detector columns increases the background
count rate in the source extraction region by a factor $\sim40$
wrt the standard imaging modes. In the MOS cameras, taking
advantage of 2-D spatial resolution, a less stringent GTI
filtering is required. GTI files were built for each of the three
cameras in order to reject the time intervals most affected by
soft protons (for MOS cameras we set thresholds of 800 cts/bin
on 40 s bin light curves extracted from peripherals CCDs in
0.2-12 keV; for the pn of 140 cts/bin on a 40 s bin light curve,
0.5-12 keV range).

\end{itemize}

The cleaned event lists correspond to dead-time corrected
exposures of 20450 s, 36900 s and 33700 s for the pn, MOS1
and MOS2 respectively. The target was detected with a 0.2-3
keV net count rate of $(2.9\pm0.3)\times 10^{-2}$ cts s$^{-1}$,
 ($3.9 \pm 0.4) \times 10^{-3}$  cts s$^{-1}$ and $(3.7\pm0.5)
\times 10^{-3}$ cts s$^{-1}$, in the pn, MOS1 and MOS2. These count
rates were estimated using a 15$''$ radius region (enclosing
$>$65\% of the source counts) in each of the MOS cameras, and a 8 pixel
wide strip (4.1$''$ pixel size) in the pn.

\section{Spectral analysis}

The source spectrum  was extracted from the regions described
above, and appropriate background regions were selected on the
same CCD chip where the target is imaged. Spectra were binned
in order to have at least 25 counts per channel for the MOSs,
80 counts per channel in the pn. {\em Ad hoc} response matrices
and effective area files were generated with the SAS tasks
{\em rmfgen} and {\em arfgen}. The spectral analysis was
performed using XSPEC v11.3. Due to the very soft spectrum of
PSR J0737-3039, the background dominates above 2.5 keV. Therefore we
restricted the analysis to the 0.2-3 keV range, which results
in 160 MOS1 counts, 130 MOS2 counts, and 2910 pn counts.
We estimate that the background accounts for $\sim$20\% of the
above values in the case of the two MOS and for $\sim$80\% in
the pn.

The spectra from MOS1, MOS2 and pn were fitted simultaneously.
An absorbed  power law model with photon index
$\Gamma$=3.5$^{+0.5} _{-0.3}$ and absorption N$_H=(7.0 \pm 2.5)
\times 10^{20}$ cm$^{-2}$, describes well the data\footnote{All the
reported errors on the spectral parameters are at 90\% confidence
for a single interesting parameter}. A confidence contour plot
for the spectral parameters  N$_H$ and $\Gamma$ is presented in
Fig.\ref{confcont}. A photon index of 2 is ruled out at more than
99.99\% level. Assuming a distance of 500 pc, the inferred 0.2-10
keV luminosity is  $\sim3.4 \times 10^{30}$ erg s$^{-1}$. An
equally good fit is obtained with a thermal bremsstrahlung model,
with temperature of $0.5 \pm 0.1$ keV and 
N$_H<5\times10^{20}$ cm$^{-2}$. Using a blackbody model the fit
is somewhat worse, with a best fit temperature of $0.16 \pm 0.01$
keV and a lower interstellar absorption
(N$_H<1.5\times 10^{20}$ ~ cm$^{-2}$). All the spectral results
are summarized in Table~\ref{spectra}. We tried also a composite
model consisting of a blackbody plus a power law with photon index
fixed at 2. The quality of the fit did not improve compared to the
best single-component models.
The 0.2-10 keV flux of a possible hard non-thermal spectral component
is $\sim2.5\times10^{-14}$  erg cm$^{-2}$  s$^{-1}$, corresponding
to a luminosity of $\sim7\times10^{29}$erg s$^{-1}$.

\section{Timing analysis}

As mentioned above, a relatively high and time variable background
affected this observation. Therefore to search for orbital
modulations in  \psr we first computed background subtracted light
curves in 10 s bins for the  MOS1 and MOS2.
For the source extraction we used a circle of radius 17$''$, while
the background was estimated from much larger source free regions
in the same CCD chips. The resulting 0.3-2 keV light curves were
then folded at the orbital period (2.45 hours), trying different
number of bins and
also varying the phases of the bin boundaries.
 We show in Fig.\ref{lcorb} the 
light curves for the MOS1, MOS2 and their sum folded in 20 bins,
as an example. No significant evidence for variability was found,
also repeating the analysis with more stringent data cleaning
selections.
Considering that the adopted extraction region contains 557 counts
($\sim$185 of which we estimate are due to the background), a 99\%
confidence level upper limit of the order of 40\% can be set on the
pulsed fraction in the simplest hypothesis of a sinusoidal modulation.

For the search of pulsations at the spin period of pulsar A we used
the pn data in the 0.3-2 keV range extracted from time intervals
with low particle background. 
This resulted in a net exposure of 13800 sec, yielding 850 counts
($\sim62$\% of which from the background). The times of arrival were
converted to the Solar system barycenter and we also corrected them
for the orbital motion and relativistic effects of the binary system
according to Blandford \& Teukolsky (1976).
This was done with a program we wrote {\em ad hoc} and tested using
a \xmm observation of the binary millisecond pulsar XTE J1751-30
(Miller et al. 2003).

Figure \ref{lcpsrA} shows the pn data folded at the expected period
of pulsar A, based on the ephemerids reported by Lyne et al. (2004).
The light curve is consistent with a constant flux. In the assumption
of a sinusoidal modulation we can set a 99\% c.l. upper limit of
$\sim60$\% on the pulsed flux from pulsar A.

Finally, we also searched for periodic modulations at the period of
pulsar B using both the pn and MOS2 data with negative results 
($\sim60$\% upper limit).

\section{Discussion}

Rotation-powered pulsars emit of the order of 10$^{-3}$
$\dot{E}$$_{rot}$ in soft X-rays (Becker \& Tr\"umper 1997;
Possenti et al. 2002). This implies that the expected luminosity
for pulsar A should be of the order of 10$^{31}$ erg s$^{-1}$ while B
should be undetectable. With a pure power-law fit of the data,
the efficiency for conversion of rotational energy of A to
X--rays is 6$\times$10$^{-4}$ (d/0.5 kpc)$^{2}$ a value
comparable with the \cha result and compatible (taking into account
a factor of $\sim$2 distance uncertainty) with the emission
originating solely from the magnetosphere of pulsar A.

The very soft observed X-ray spectrum ($\Gamma$=3.5) is not
unheard of when considering the class of recycled pulsars.
Such sources are known to show broadly different spectral
phenomenologies (see e.g. Becker \& Aschenbach 2002 and
references therein; Webb et al. 2004a,b), ranging from very
hard, non-thermal spectra with power law photon index
$\Gamma<2$ (e.g. PSR B1821-24, PSR B1937-21), to very soft
($\Gamma \sim 3-4$) and possibly thermal spectra (e.g.
PSR J0030+0451, PSR J2124-3358). Indeed, PSR J0437-4715,
the nearest and brightest ms pulsar, was already known in
the ROSAT era to have a composite spectrum showing both a
soft and a hard component (Zavlin \& Pavlov 1998).
Thus, an interpretation of the observed X-ray emission
from the \psr system as ``normal'' recycled pulsar
radiation originating from pulsar A seems the simplest
one. The upper limit on the pulsed fraction at the 22 ms
period ($\sim60$\%) is not inconsistent with this picture,
since modulations at the $\sim30-40$\% level have been
observed in some cases.

However, the \psr system offers other possible sources of
high-energy photons with comparable power than magnetospheric
emission from A. A remarkable one might be the bow-shock
that forms near pulsar B owing to the collision between A's
relativistic wind and B's magnetosphere (Arons et al. 2004).

If the magnetisation of A's wind is low, the shock is
expected to be strongly dissipative and a large fraction
of the incoming energy flux may be radiated in X-rays.
Within distance uncertainty, the observed luminosity of
$\sim$3$\times$10$^{30}$ (d/0.5 kpc)$^{2}$ erg s$^{-1}$ is
compatible with a fraction ($\le$50\%) of the maximum
X--ray flux expected from the bow-shock.
This limit is set by
L$_{max}$$\sim$$\dot{E}$$_{A}$$\Delta$$\Omega$$/4\pi$$\sim$3$\times$10$^{31}$
erg s$^{-1}$, where $\Delta$$\Omega$$\sim$6$\times$10$^{-2}$sr 
(Lyutikov 2004) is the solid angle seen from pulsar A
enclosing the magnetosheath around pulsar B.

As already seen with {\textit{Chandra}}, the X--ray photon index  is
rather steep, at variance with the flat $\nu$F$_{\nu}$ spectrum typically expected
from shock acceleration.
This is independently confirmed by our data
which give a range of allowed slopes even less compatible with the
``canonical'' shock value $\Gamma\sim$2  (see Fig.1).
Such a soft spectrum could be due to the presence
of an unusual low-energy relativistic electron population as
required in the magnetosheath model (Arons et al. 2004; Lyutikov 2004)
to provide A and B's eclipses by synchrotron absorption.

Within the bow-shock/magnetosheath model, one might expect a
flux modulation as a function of the orbital period owing to
the changing view of the shock front. In particular, the shocked
wind is expected to flow away from the head of the bow-shock in
a direction roughly parallel to the shock (Lyutikov 2004; Granot \& Meszaros 2004).
This might imply relativistic beaming of the radiation emitted
by the shocked plasma, resulting in a $\lesssim$50$\%$ modulation
of the observed emission as a function of the orbital phase
with possible peaks at 90 degrees from conjunctions.
Our analysis of the flux variations as a function of the
orbital phase allow us to exclude the presence of any
large orbital modulation $\ge$40$\%$.
However, we note that a low significance peak (MOS1 data)
is seen in the orbital light curve shifted roughly 0.3
in phase from inferior conjunction.

It is worth considering also possible modulation of the X--ray
flux as a function of the spin period of pulsar B (2.77 s). The rotation of
pulsar B, assuming a misalignement of its magnetic pole relative
to its spin axis (Ramachandran et al. 2004), would change the
magnetic pressure behind the bow-shock as well as the values of
the physical quantities of the shocked plasma with a periodicity
equal to the spin period. No periodicity with
P$_{B}$ was detected in pn and MOS data.

Although not improving the fit to the data with respect to
one-component models, the scenarios involving a non-thermal
component, originating from A's magnetosphere and/or from a
bow-shock, superimposed to a thermal one is still an
intriguing possibility. 
The luminosity of the non-thermal
component
we derived assuming a power-law with canonical shock acceleration photon index $\sim$2
 is nearly an order of magnitude smaller than that
expected from the bow-shock/magnetosheath model
(Granot \& Meszaros 2004).
These authors suggest that the likeliest explanation for the
X--ray emission is the pulsar A wind just behind the bow-shock
caused by the systemic motion in the ISM.
This model provides luminosities of $\sim$6$\times$$10^{29}$ erg s$^{-1}$,
a value in agreement with our results for the composite BB+PL fit.
In this case, little or no modulation with the orbital period,
P$_{A}$ and P$_{B}$ is expected.

The remarkable steepness of the spectrum might be related to
blackbody emission with temperature $\sim$0.15 keV and
luminosity roughly half of the total X-ray luminosity.
Thermal emission from polar caps could still play a role for
pulsar A, despite its large spin-down age ($\sim$210 Myrs).
Furthermore, it is in principle conceivable that a portion of A's
relativistic Poynting and particle flux might heat the
surface of B, giving rise to thermal radiation (Zhang \& Loeb 2004). To match
both luminosity and black-body temperature, one could consider
the possibility that part of A's wind energy is absorbed by
B's magnetosphere and driven towards B surface. If the expected
energy input from A ($\sim$10$^{31}$ erg s$^{-1}$) could be transferred
with an efficiency $\ge$10\%, at B's surface, the resulting
heating could match the thermal luminosity of $\sim$10$^{30}$
erg s$^{-1}$ obtained from our spectral fits.
There is no way to obtain high temperatures ($>$1 eV) directly
from the heating of structures with magnetospheric size.

Future high-energy observations of \psr could better trade
spectral models and reduce upper limits on orbital modulation
and pulsed fractions from A and B providing conclusive evidence 
of the magnetospheric, thermal or bow-shock origin of the X--ray flux.
For example, another $\sim$50 ks \xmm/EPIC observation with pn 
operating in ``Small Window'' mode, trading sensitivity vs.
time resolution, could strongly improve the search for P$_{orb}$
and P$_{B}$ modulations as well as spectral fits, while further 
observations with pn in ``Fast Timing'' mode, looking for A's
pulsations, should be much longer if aimed to detect the
small pulsed fraction in principle expected from a relatively
weak millisecond pulsar.
~\\

Based on observations with {\textit{XMM-Newton}}, an ESA science
mission with instruments and contributions directly
funded by ESA member states and the USA (NASA).
We thank Norbert Schartel for approving this observation within
the frame of discretionary time of Project Scientist.

\clearpage

\begin{table}[!h]
\begin{center}
\caption{\label{spectra} Results of the spectral analysis. The quoted errors
  are at 90\% confidence for a single interesting parameter.
The errors on fluxes are at 1$\sigma$ level.}
\vspace{1cm}
\begin{tabular}{lcccc}
\hline \hline
 & Power Law & Bremsstrahlung & Blackbody & BB + PL \\

 \hline

 \hline
$\chi^2_{\nu}$/d.o.f. & 0.92$/$40 & 0.86$/$40  & 1.11$/$40  & 0.90$/$39 \\

\hline

N$_H$ (cm$^{-2}$) & 7.0$\pm$2.5 & $<$5 & $<1.5$ & $<4$ \\

\hline

$\alpha_{ph}$ & 3.5$^{+0.5} _{-0.3}$ & - & - & 2 (fixed) \\

\hline

kT (keV) & - & 0.5$\pm$0.1 & $0.15 \pm 0.01$ & $0.14 \pm 0.01$ \\

\hline

F$_{0.2-3 keV}^{(a)}$ (10$^{-14}$ erg cm$^{-2} $ s$^{-1}$) &
$3.7\pm0.5$ & $4.1\pm0.5 $  & $3.3\pm0.5$  & $4.0\pm0.5$ \\

\hline

 L$_{0.2-10 keV}^{(b)}$ (erg s$^{-1}$) & $3.0 \times
10^{30}$  & $1.1 \times
 10^{30}$ & - & $6.9 \times 10^{29}$ \\

 \hline

L$_{BB}^{(c)}$ (erg s$^{-1}$)   & - & - & $9.8\times 10^{29}$ &
$7.1\times
 10^{29}$ \\

 \hline

 \hline
\end{tabular}
\end{center}
$^a$ Observed flux in the 0.2-3 keV range. \\
$^b$ 0.2-10 keV luminosity (excluding BB component) for a distance of 500 pc. \\
$^c$ bolometric luminosity for BB component for a distance of 500 pc.\\
\end{table}

\clearpage

\figcaption{Contour plot of spectral parameters for the absorbed
power law model. The plotted confidence levels are 68.3\%, 90\%
and 99\%.\label{confcont}}

\figcaption{Background subtracted light curves folded at the
orbital period (8834.5350432 s). MOS1 (top), MOS2 (middle) and MOS1+MOS2 (bottom).
The corresponding $\chi^{2}$ values are 17.9 (MOS1), 10.8 (MOS2) and 
24.3 (MOS1+MOS2) for 19 degrees of freedom.
The dashed line marks the inferior conjunction of B (phase=0.43)
when the two stars are at their closest on the sky and B is nearer
to the Earth.
 \label{lcorb}}

\figcaption{PN data folded at the period (P=22.6993786 ms) expected for
pulsar A \label{lcpsrA}. The dashed line indicates the estimated
background contribution which has not been subtracted since it does 
not vary on this short time scale.}


\clearpage
\epsscale{.7}
\plotone{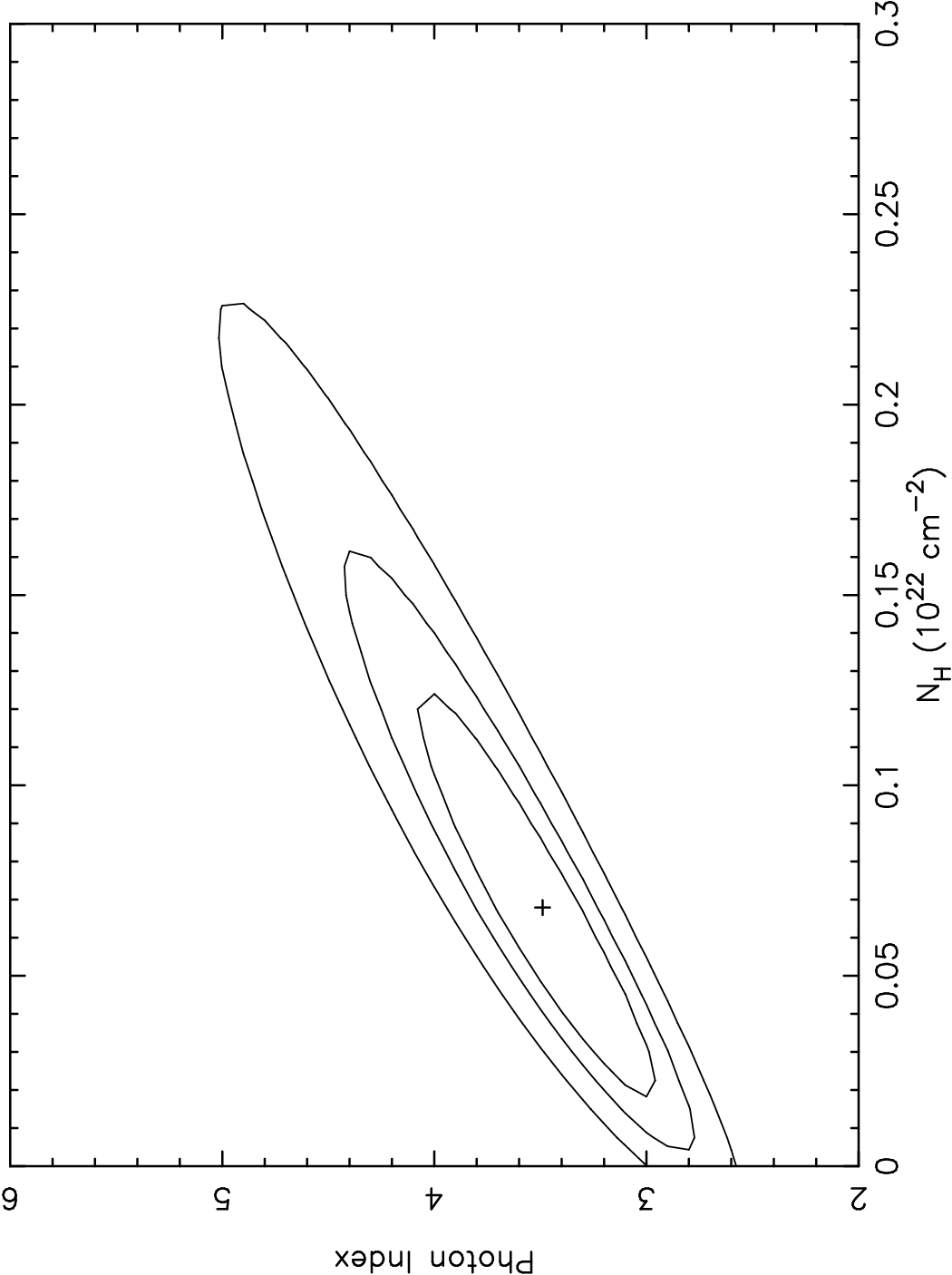}

\clearpage
\epsscale{.7}
\plotone{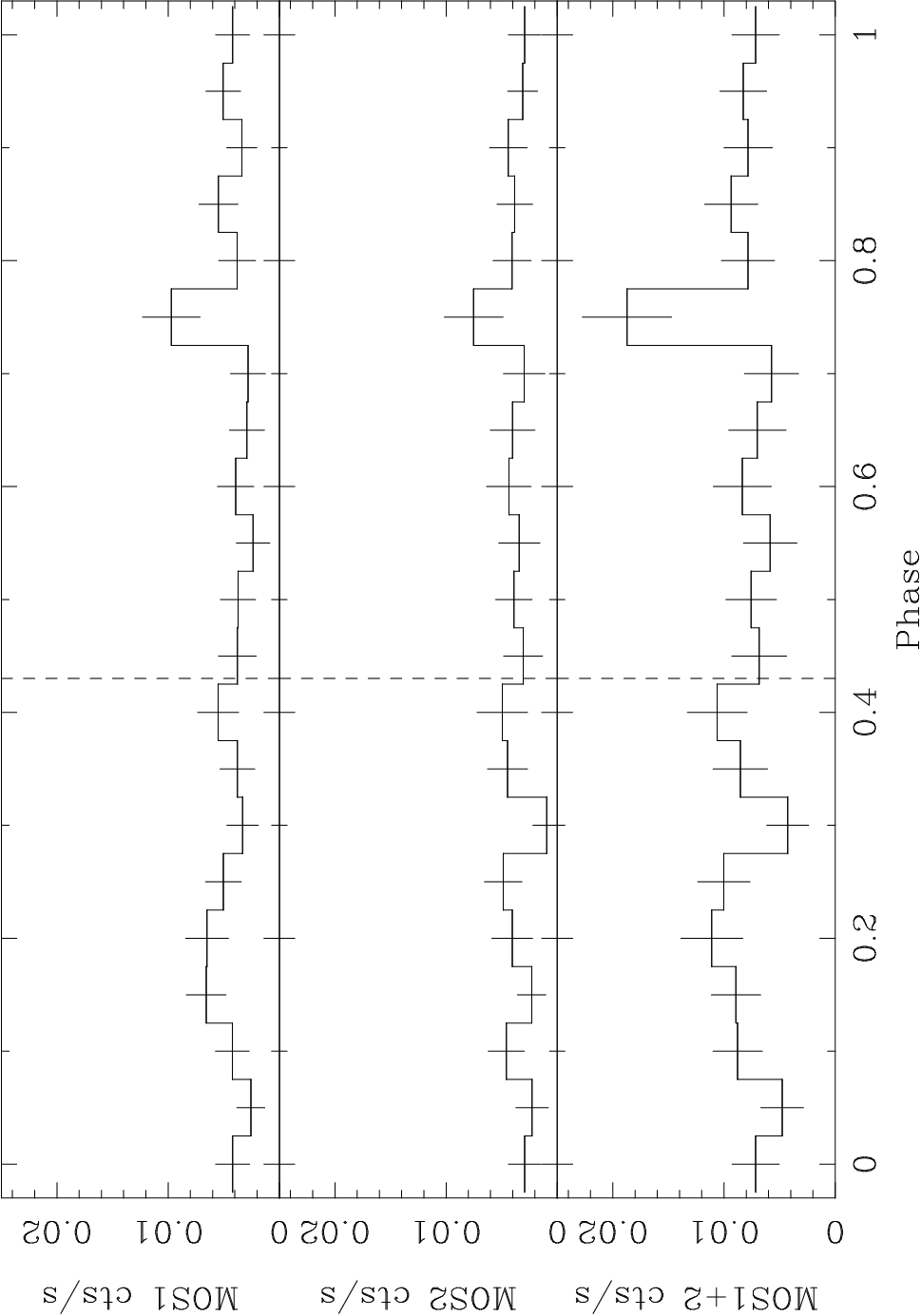}

\clearpage

\epsscale{.7}

\plotone{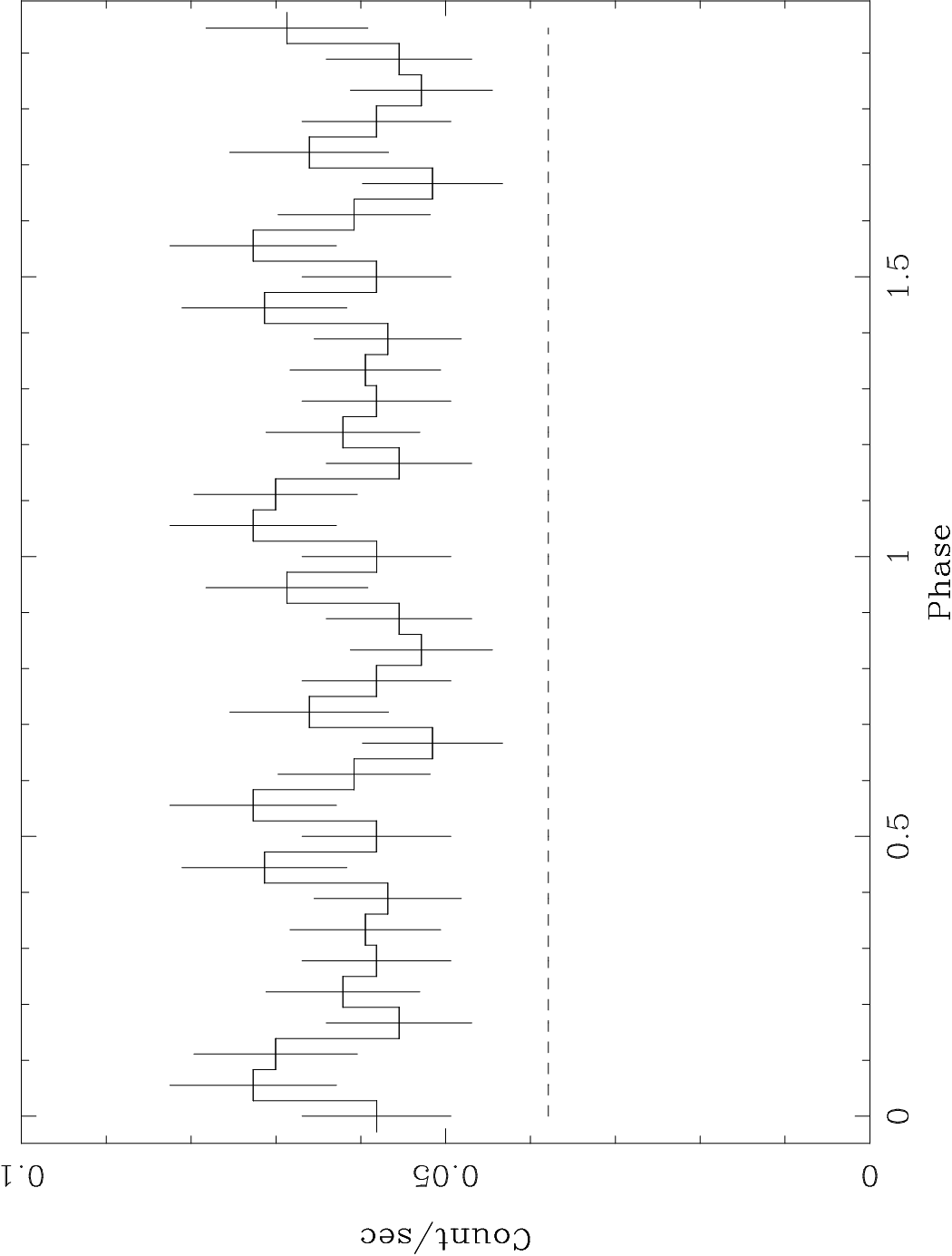}

\end{document}